\documentclass[iop]{emulateapj}
\usepackage{epsfig}

	\newcommand{\chandra}{{\it Chandra}}

	\newcommand{\rxte}{{\it RXTE}}
	\newcommand{\xmm}{{\it XMM-Newton}}
	\newcommand{\swift}{{\it Swift}}
	\newcommand{\maxi}{MAXI\,J1659--152}

 \def\spose#1{\hbox to 0pt{#1\hss}}
\def\laeq{\mathrel{\spose{\lower 3pt\hbox{$\mathchar"218$}}
     \raise 2.0pt\hbox{$\mathchar"13C$}}}
\def\gaeq{\mathrel{\spose{\lower 3pt\hbox{$\mathchar"218$}}
     \raise 2.0pt\hbox{$\mathchar"13E$}}}

\begin{document}

\slugcomment{Accepted for publication in ApJL}

	\title{Optical emission of the black hole X-ray transient MAXI J1659--152 during quiescence}
	\author{Albert~K.~H.~Kong\altaffilmark{1}}
	\affil{Institute of Astronomy and Department of Physics, National Tsing Hua University, Hsinchu 30013, Taiwan; akong@phys.nthu.edu.tw}
	\altaffiltext{1}{Golden Jade Fellow of Kenda Foundation, Taiwan}

\begin{abstract}
We report on the optical detection of the black hole X-ray transient \maxi\ during its quiescent state. By using the Canada France Hawaii Telescope (CFHT), we observed \maxi\ about 7 months after the end of an X-ray outburst. The optical counterpart of \maxi\ is clearly detected with a $r'$-band magnitude of 23.6--23.8. The detection confirms that the optical emission of \maxi\ during quiescence is relatively bright comparing to other black hole X-ray transients. This implies that the distance to \maxi\ is 4.6--7.5 kpc for a M2 dwarf companion star, or 2.3--3.8 kpc for a M5 dwarf companion star. By comparing with other measurements, a M2 dwarf companion is more likely.

\end{abstract}


\section{Introduction}
Galactic black hole X-ray binary systems manifest themselves as X-ray novae and most of them are discovered via their sudden dramatic increase in X-ray brightness. Such an X-ray outburst is due to a change of the amount of material accreted from the companion star by the central compact object. In addition to X-ray follow-up observations following the evolution of the outburst, multi-wavelength observations play an important role in understanding the physics of these systems. In particular, reprocessing in the accretion disk and the companion star will also generate optical emission. We therefore expect to see a dramatic change in optical brightness when an X-ray nova is in outburst. 

Optical observations provide the data necessary to determine the orbital period, the evolutionary state of the secondary, the binary mass function, and mass of the compact object itself. In combination with X-ray spectral and timing studies, these provide a detailed picture of the accretion process and the nature of the compact object. Although the optical counterpart of an X-ray nova is usually discovered during an outburst, observations when the source returns to quiescence are also extremely important. While the optical emission of an X-ray nova is very faint in quiescence, it is the best time to study the nature of the companion star and the geometry of the binary system, and to measure the mass function via radial velocity measurements.

The X-ray transient \maxi\ was first discovered with the \swift\ Burst Alert Telescope (BAT) on 2010 September 25 as a gamma-ray burst (GRB 100925A; Mangano et al. 2010). Independent discovery made with the Gas Slit Camera (GSC; Mihara et al. 2011) on board the {\it Monitor of All-sky X-ray Image} ({\it MAXI}; Matsuoka et al. 2009) suggested that it is a previously unknown Galactic X-ray transient. The optical counterpart of \maxi\ was discovered with the \swift\ UltraViolet/Optical Telescope (UVOT) immediately after the BAT trigger (Marshall 2010). Subsequent multi-wavelength observations from radio to hard X-ray showed that \maxi\ is a black hole binary candidate. In particular, frequent X-ray dips with a recurrent time of 2.4 hours were found in \rxte, \swift, and \xmm\ observations (Kennea et al. 2011; Kuulkers et al. 2012). This strongly suggests that \maxi\ is a highly inclined system with an orbital period of 2.4 hours, the shortest among all the black hole X-ray binaries.

\maxi\ returned to quiescence as observed with \chandra\ on 2011 May 3 (Jonker et al. 2012), but it underwent a mini-outburst from 2011 May 6 (Yang \& Wijnands 2011a,2011b; Jonker et al. 2012). The source has finally settled down in the quiescent state from 2011 mid-August based on a \chandra\ monitoring program (Jonker et al. 2012).

The optical counterpart of \maxi\ has been monitored regularly with \swift\ UVOT and ground-based telescopes during the outburst (Russell et al. 2010; Yang et al. 2011; Yang \& Wijnands 2011a,2011b). In general, the optical activity tracked the X-ray outburst (Russell et al. 2010,2011; Kong et al. 2011). The optical counterpart of \maxi\ is not detected in the Digitized Sky Survey indicating that the quiescent magnitude is $> 21$. A tentative quiescent counterpart was reported by Kong et al. (2010) using the Pan-STARRS 1 (PS1) data. The source was only marginally seen in the image with a $r'$-band magnitude of $\sim 22.4$. This was challenged by Kennea et al. (2011) and Kuulkers et al. (2012) noting that it is too bright for \maxi.

In this paper, we report our new deep optical imaging observation of \maxi\ in quiescence using the Canada France Hawaii Telescope (CFHT) aimed at measuring an accurate quiescent optical magnitude. We also re-analysed the PS1 data for comparison.

\begin{figure*}
	\centering
	\psfig{file=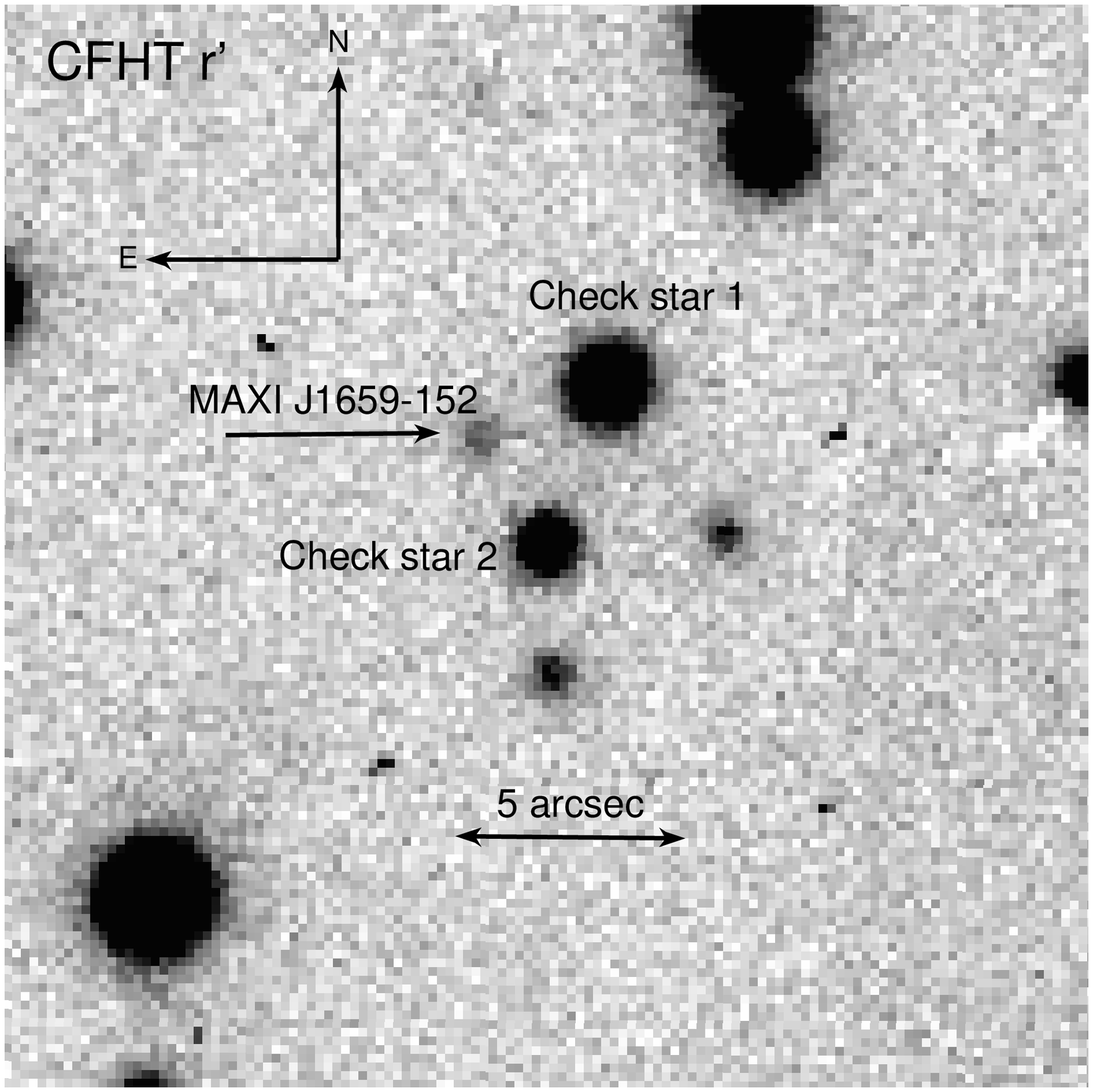,width=59mm}
	\psfig{file=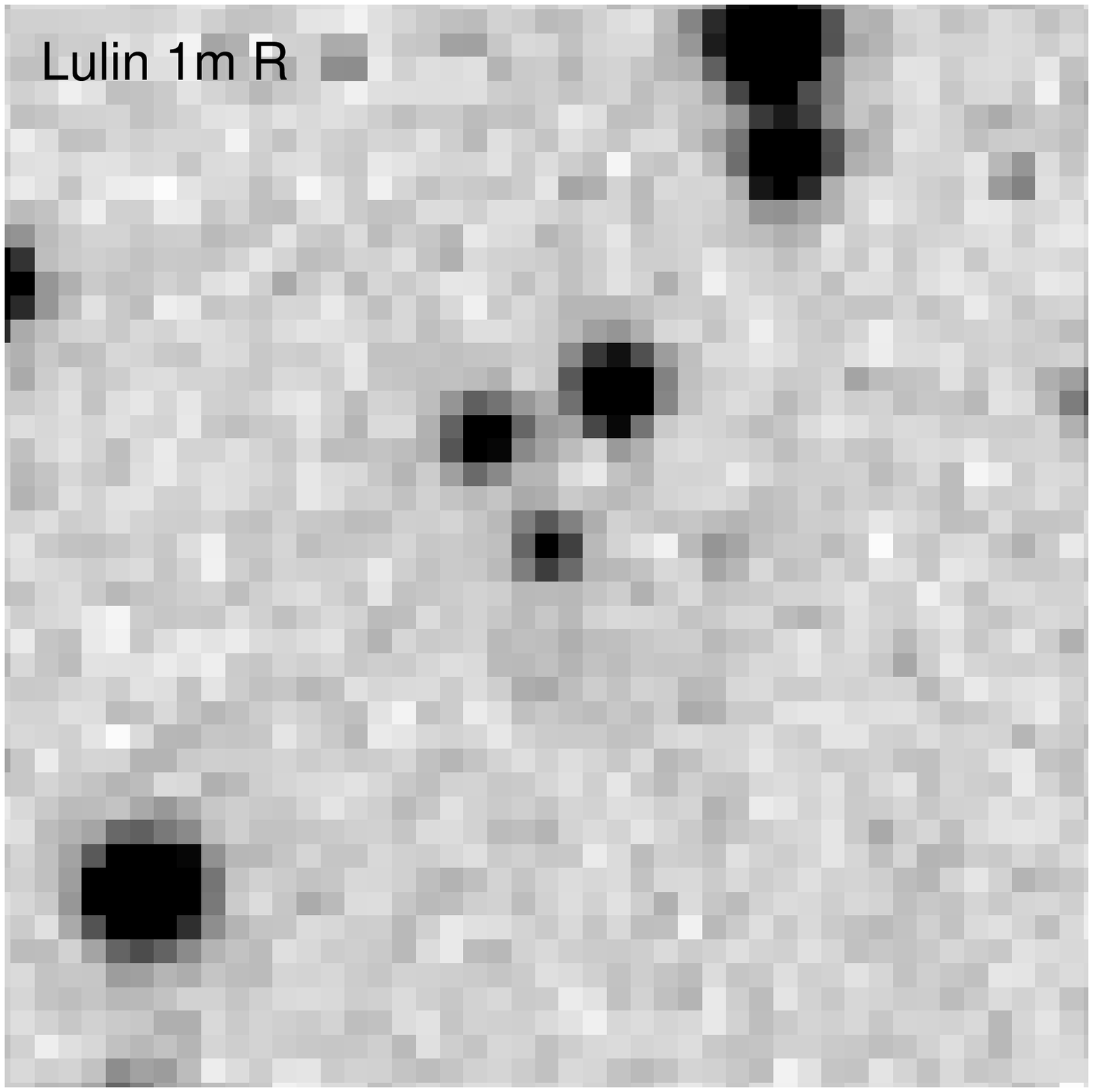,width=59mm}
	\psfig{file=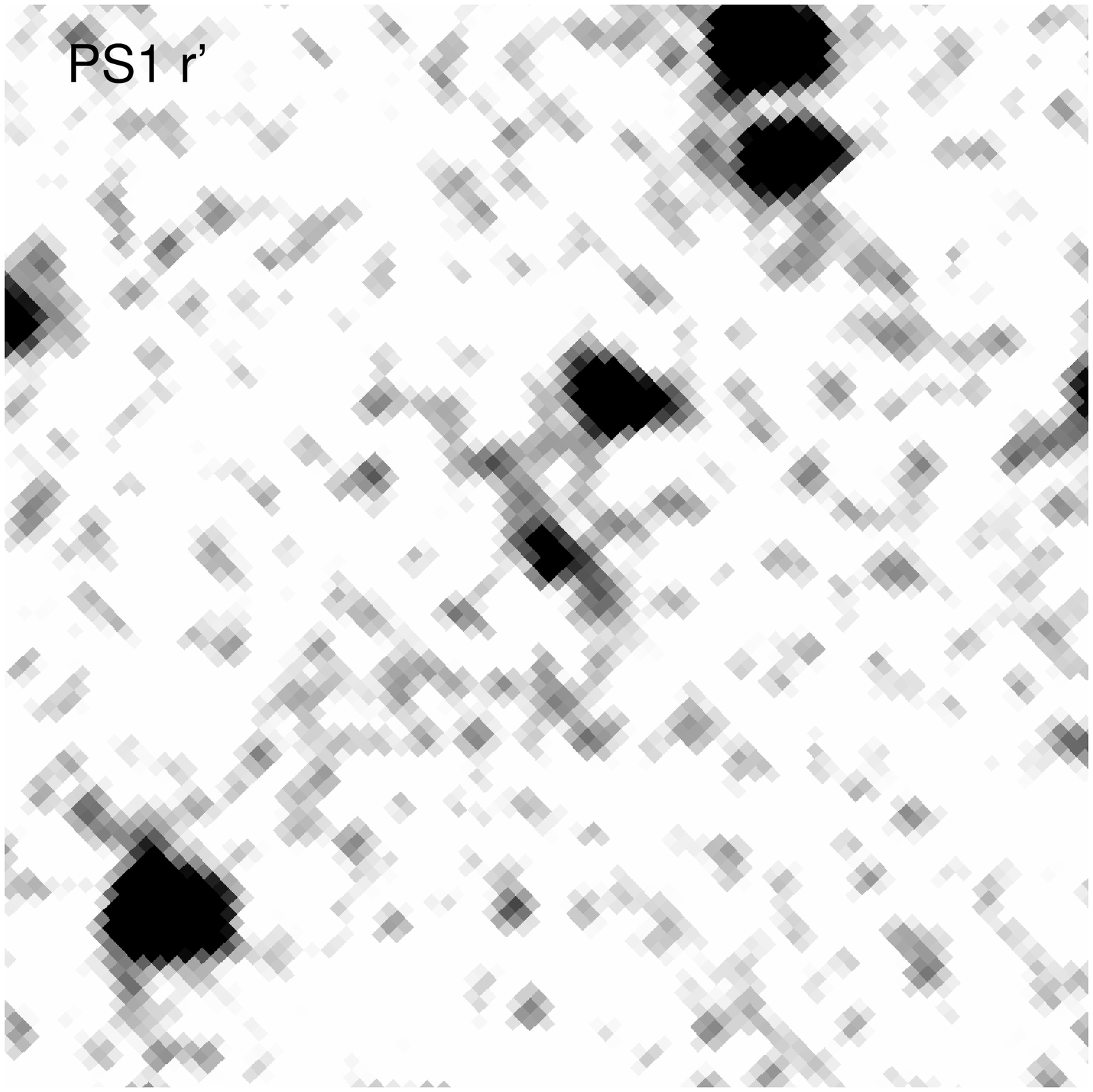,width=59mm}
	\caption{CFHT/MegaCam $r'$-band (left), Lulin 1m $R$-band (middle), and PS1 $r'$-band (right) images of \maxi. The CFHT image was taken on 2012 March 23, about 7 months after the end of an X-ray outburst. \maxi\ is clearly seen. The Lulin 1m image was taken with the 1m telescope at the Lulin Observatory in Taiwan; the image was taken near the end of an X-ray outburst on 2011 July 30 and the source has a $R$ magnitude of $\sim 21$ (see also Kong et al. 2011). Comparing with the CFHT image, \maxi\ is much brighter in the Lulin image. The PS1 image was taken as part of the 3$\pi$ sky survey on 2010 June 19, 3 months before the discovery of an X-ray outburst (Kong et al. 2010).}
\end{figure*}

\section{Observations and Data Analysis}
\subsection{CFHT}
\maxi\ was observed with the MegaPrime/MegaCam at the CFHT on 2012 March 23. We checked the {\it MAXI} GSC daily data and the source was not detected in soft X-ray indicating that it was likely in quiescence. However, we caution that if there is an X-ray mini-outburst, \maxi\ will not be detected with {\it MAXI} due to its low sensitivity. 
The MegaCam has an array of 36 CCDs, giving a total of 1 degree by 1 degree field-of-view. We obtained two $r'$-band images under a seeing condition of $\sim 0\farcs8$ with an exposure time of 980 seconds each. The two images were separated by about 30 minutes. The raw images were processed with Elixir\footnote{http://www.cfht.hawaii.edu/Instruments/Elixir} for bias, flat-field, overscan, bad pixel mask, and zero point.

We next corrected the CFHT images for the astrometry. By using 10 stars surrounding \maxi, we used the IRAF task {\it ccmap} to compare with the USNO point source catalog. The resulting registration errors are $0\farcs21$ in R.A. and $0\farcs226$ in declination.

By examining the CFHT image together with images taken during the outburst (Russell et al. 2010; Kong et al. 2011; see also Fig. 1), the optical counterpart of \maxi\ is clearly a variable (see Figure 1). We further compared the CFHT position of \maxi\ with the European VLBI Network position (Paragi et al. 2010); the offset between the two positions is $0\farcs043$ indicating that it is the true optical counterpart. We then performed aperture photometry using the task {\it phot} in IRAF. In addition to our target, we also measured the optical magnitude for a few nearby stars for checking (see Fig. 1). 

\subsection{PS1}
We also re-analysed the PS1 image taken before the outburst of \maxi\ (Kong et al. 2010) by using the same method for the CFHT data. We obtained postage stamp images from the Postage Stamp Server maintained by the PS1 Image Processing Pipeline team at the University of Hawaii. As part of the $3\pi$ survey, the field of \maxi\ was observed with various filters several times each, but the source was only marginally seen with the $r'$-band filter taken on 2010 June 19 with an exposure time of 40 seconds. Since it is a few months before the first discovery of the outburst, the source should be in quiescence. As a check, the {\it MAXI} GSC light curve did not show any enhancement in the X-ray flux. Since \maxi\ is not in the $3\pi$ survey catalog, we determined its magnitude by comparing with several catalogued stars. Like the CFHT data, we performed an aperture photometry for \maxi\ as well as the comparison and check stars. The magnitudes of the check stars measured by us are entirely consistent with the $3\pi$ survey catalog.

\section{Results}

In Table 1, we list the $r'$-band magnitudes of \maxi\ and a couple of check stars of the CFHT and PS1 observations. Given that our new CFHT observations were taken more than 7 months after the end of an X-ray outburst confirmed by \chandra\ (Jonker et al. 2012) and there is no unusual X-ray activity shown in the {\it MAXI} GSC data, \maxi\ is likely in the quiescent state although we cannot totally rule out a mini-outburst or a very long optical decay. The $r'$-band magnitude of \maxi\ varied between $23.62\pm0.06$ and $23.81\pm0.06$ in 30 minutes. The two check stars near \maxi\ did not show any variability, indicating that \maxi\ is likely a variable star.

For the PS1 data, \maxi\ is marginally visible with a $r'$-band magnitude of $22.8\pm0.3$ (see Fig. 1). This is roughly consistent with the quick-look analysis reported by Kong et al. (2010). We examined all the PS1 images in this field and the observation reported here has the best seeing and image quality among all the images. This may explain why \maxi\ was only detected in this observation. Furthermore, the CFHT images indicate that \maxi\ can be fainter than the PS1 measurement. It could happen that the PS1 observation was taken during a relatively bright state in the quiescence.

\begin{table*}
	\centering
	\footnotesize
\caption{Optical ($r'$-band) magnitudes of \maxi\ and check stars}
\begin{tabular}{cccc}
	\hline
	\hline
Source &	\multicolumn{2}{c}{CFHT} & PS1\\
	\cline{2-3}
 & Epoch 1 & Epoch 2 &  \\
\hline
\maxi\ & $23.81\pm0.06$  & $23.62\pm0.06$ & $22.8\pm0.3$\\
Check star 1 &  $20.738\pm0.004$ & $20.731\pm0.005$ & $20.65\pm0.05$\\
Check star 2 &  $21.85\pm0.01$ & $21.86\pm0.01$ & $21.7\pm0.1$ \\
\hline
\hline\\
\end{tabular}
\end{table*}

\section{Discussion}

From both the CFHT and PS1 observations, we can conclude that the optical counterpart of \maxi\ shows variability during or near quiescence. In Figure 1, we also show a $R$-band image taken near the end of an outburst with the 1m telescope at the Lulin Observatory in Taiwan (Kong et al. 2011). \maxi\ is clearly much fainter during our CFHT observations. Optical variability is quite common for black hole  X-ray transients in quiescence. For example, GRS\,1124-684, A0620-00, J0422+32, GS\,2000+25, and V404 Cyg show flarings on timescales of a few minutes to 60 minutes with a typical amplitude of 0.1--0.6 mag (Zurita et al. 2003; Shahbaz et al. 2003,2010). BW Cir (=GS\,1354-64) is another quiescent black hole that exhibits large (0.5--1 mag) optical variability (Casares et al. 2009). The origin of optical flaring during quiescence is not well understood. Possible mechanisms include X-ray irradiation of the accretion disk (Hynes et al. 2004), magnetic reconnection events (Zurita et al. 2003), and  direct synchrotron emission from an advective dominated flow (Shahbaz et al. 2003). Simultaneous X-ray and optical observations are required to understand their nature (e.g., Hynes et al. 2004).

In addition to optical flaring activity, the orbital variation can contribute to the observed variability. \maxi\ is shown to have a 2.4-hour orbital period based on the dip-like events in the X-ray light curves (Kuulkers et al. 2012). In particular, the two CFHT observations are separated by about 30 minutes, equivalent to about 0.2 orbital phase. Variability is therefore not unexpected. The 0.2 mag difference is also consistent with other black hole X-ray transients with measured optical orbital modulation (e.g., see Fig. 1 of Zurita et al. 2003). Future optical monitoring observations will be crucial to reveal the optical orbital modulation. 

Our new CFHT observations strongly suggest that the proposed optical counterpart of \maxi\ is real and it shows noticeable variability that could be from flaring or orbital modulation. Even we adopt the faintest measurement from our CFHT observations ($r'=23.8$), it is in contrast to the suggestion that the quiescent magnitude is $> 27$ (Kuulkers et al. 2012). We note that the orbital modulation cannot produce such a huge optical variability ($> 3$ mag) unless the compact object has strong radiation to evaporate the companion like black widow-type pulsars (e.g., Kong et al. 2012). The suggested quiescent optical magnitude is based on a relationship between the optical outburst amplitude of an X-ray transient and the orbital period found by Shahbaz and Kuulkers (1998; hereafter SK98). With an orbital period of 2.4 hours, the $V$-band magnitude difference between the outburst and quiescence should be as large as 11 magnitudes, yielding $V > 27$ in the quiescence. Taking the inclination effect into account, the quiescent $V$ magnitude is $> 26.2$ (Kuulkers et al. 2012). By using previous PS1 measurement in the $r'$ band (Kong et al. 2010) and the limiting magnitudes of the PS1 $3\pi$ survey, Kuulkers et al. (2012) derived an expected $V$-band magnitude of $V\gaeq22.8$; although it cannot rule out a much fainter object, a difference of at least 3 magnitudes indicates that \maxi\ unlikely follows the SK98 relation.  However, we caution that the SK98 relation is an empirical relation based on a sample of 11 black hole transients and there are many factors (e.g., the stellar type of the companion star, the inclination effect on the disk emission, and the peak mass transfer rate) affecting the correlation. Hence, scatter on the correlation is not unexpected.

To estimate the $V$-band magnitude, we first assume the spectrum of the companion. Given that \maxi\ is likely a M2 or M5 dwarf (Kuulkers et al. 2012; Jonker et al. 2012), the colors will become $B-V=1.52$ (M2 dwarf) or $B-V=1.61$ (M5 dwarf). We then correct the colors for the reddening assuming the Galactic values provided by Schlafly et al. (2011).
Using the transformation of Tonry et al. (2012), we get a reddened $r'-R$ color of 0.3 for both cases. The reddened $V-R$ is 1.9 and 2.2 for a M2 dwarf and a M5 dwarf, respectively. This implies that $V-r'$ is 1.6 (1.9) for a M2 dwarf (M5 dwarf). Using the observed $r'$-band magnitude (23.8) from our CFHT observation, we have $V=25.4$ for a M2 dwarf and $V=25.7$ for a M5 dwarf. These are consistent with the above limits derived by Kuulkers (2012). In any case, the estimated quiescent $V$-band magnitude is within 1 magnitude difference for a high-inclination ($80^{\rm o}$) system although the difference could be as large as 2 magnitudes for a lower inclination ($65^{\rm o}$) system. 

Based on our new CFHT measurement, although the estimated $V$-band magnitude is still brighter, it is not a large offset as proposed by Kuulkers et al. (2012). As we mentioned above, there are other factors causing scatter on the SK98 correlation. For instance, 
the black hole X-ray binary XTE J1118+480 has an orbital period of 4.08 hours and the outburst amplitude is about 6--7 magnitudes. According to SK98, the amplitude is at least 9.4 magnitudes. However, XTE J1118+480 is underluminous in outbursts; the outburst X-ray spectra are hard in contrast to typical high/soft state of X-ray transients in outbursts, and the X-ray-to-optical flux ratio is low ($\sim 5$; see Zurita et al. 2006 for a discussion).  Unlike XTE J1118+480, the X-ray spectral behaviors and X-ray-to-optical flux ratio of \maxi\ are similar to a typical X-ray transient in outburst (Kennea et al. 2011).
Follow-up photometric and spectroscopic observations of \maxi\ in the future will be able to provide better constraints. On the other hand, the quiescent X-ray luminosity of \maxi\ as measured with \chandra\ is an order of magnitude higher than that derived from the orbital period-X-ray luminosity correlation (Jonker et al. 2012). Given that both optical and X-ray brightness are higher than expected, one simple explanation is that \maxi\ is not in a true quiescent state. It is worth noting that some short orbital period black hole systems have a long decay time. For example, GRO\,J0422+32 reached its optical quiescent state 760 days after an outburst (Garcia et al. 1996). Further optical and X-ray observations will confirm this. 

We now can use the absolute magnitude of the companion to estimate the distance to the source. The absolute $V$-band magnitude of a M2 dwarf and M5 dwarf is 10 and 11.8, respectively. Since $V-R$ is 1.5 (1.8) for a M2 (M5) dwarf, this implies $M_R$ is 8.5 (10). By using the transformation of Tonry et al. (2012), $M_r$ becomes 8.77 (10.27). Comparing with the reddening ($A_r=1.4$; Schlafly et al. 2011) corrected $r'$-band magnitude, the distance is 5.3 kpc (M2 dwarf) or 2.7 kpc (M5 dwarf). If we further assume the accretion disk contributes 50\% of the visible light (see Jonker et al. 2012), then the distance would be between 7.5 kpc (M2 dwarf) and 3.8 kpc (M5 dwarf).
If $A_r$ is as large as 1.7 (Schlegel et al. 1998), then the distance to \maxi\ will be 4.6--6.5 kpc and 2.3--3.3 kpc for a M2 dwarf and a M5 dwarf, respectively.
Comparing with previous estimations (e.g., Kennea et al. 2011; Kaur et al. 2012; Kuulkers et al. 2012; Jonker et al. 2012), a M2 dwarf companion star is more likely. This is also consistent with the conclusion made by Jonker et al. (2012). On the other hand, if \maxi\ has a M5 dwarf companion star and a distance of $\sim2$ kpc, the observed X-ray luminosity will be consistent with the orbital period-X-ray luminosity correlation (Jonker et al. 2012).
Future optical spectroscopy during quiescence will be crucial to determine the stellar type of the companion star and hence the distance to the system. In addition, an ellipsoidal lightcurve modeling during quiescence will allow us to derive the inclination angle of the system. These observations will also test if the SK98 correlation holds for \maxi. Combining with a deep X-ray observation during the quiescent state, we could constrain the orbital period-X-ray luminosity correlation.

\begin{acknowledgements}
We thank an anonymous referee for comments that have improved this paper.
Ground-based observations were obtained with MegaPrime/MegaCam, a joint project of CFHT and CEA/DAPNIA, at the Canada-France-Hawaii Telescope (CFHT) which is operated by the National Research Council (NRC) of Canada, the Institut National des Science de l'Univers of the Centre National de la Recherche Scientifique (CNRS) of France, and the University of Hawaii.	Access to the CFHT was made possible by the Institute of Astronomy and Astrophysics, Academia Sinica, National Tsing Hua University, and National Science Council, Taiwan.  

The Pan-STARRS1 Surveys (PS1) have been made possible through contributions of the Institute for Astronomy, the University of Hawaii, the Pan-STARRS Project Office, the Max-Planck Society and its participating institutes, the Max Planck Institute for Astronomy, Heidelberg and the Max Planck Institute for Extraterrestrial Physics, Garching, The Johns Hopkins University, Durham University, the University of Edinburgh, Queen's University Belfast, the Harvard-Smithsonian Center for Astrophysics, the Las Cumbres Observatory Global Telescope Network Incorporated, the National Central University of Taiwan, the Space Telescope Science Institute, and the National Aeronautics and Space Administration under Grant No. NNX08AR22G issued through the Planetary Science Division of the NASA Science Mission Directorate. We thank the PS1 Builders and PS1 operations staff for construction and operation of the PS1 system and access to the data products provided.

This project is supported by the National Science Council of the Republic of China (Taiwan) through grant NSC100-2628-M-007-002-MY3 and NSC100-2923-M-007-001-MY3. A.~K.~H.~K. gratefully acknowledges support from a Kenda Foundation Golden Jade Fellowship. 
\end{acknowledgements}

{\it Facilities:} \facility{CFHT (MegaCam)}, \facility{Pan-STARRS1}

\end{document}